\documentclass[conference]{IEEEtran}
\IEEEoverridecommandlockouts
\usepackage{cite}
\usepackage{amsmath,amssymb,amsfonts}
\usepackage{algorithmic}
\usepackage{graphicx}
\usepackage{textcomp}
\usepackage{xcolor}
\usepackage{verbatim}
\usepackage{comment}
\usepackage[colorlinks=true, linkcolor=blue, urlcolor=blue,citecolor=blue]{hyperref}
\def\BibTeX{{\rm B\kern-.05em{\sc i\kern-.025em b}\kern-.08em
    T\kern-.1667em\lower.7ex\hbox{E}\kern-.125emX}}

\begin{document}

\title{OpenNIRScap: An Open-Source, Low-Cost Wearable Near-Infrared Spectroscopy-based \\Brain Interfacing Cap
}

\author{\IEEEauthorblockN
        {Tony Kim,
        Haotian Liu,
        Chiung-Ting Huang,
        Ingrid Wu,
        and~Xilin~Liu
        }
\IEEEauthorblockA{The Edward S. Rogers Sr. Department of Electrical \& Computer Engineering (ECE),\\ University of Toronto, Toronto, ON, Canada }}

\maketitle

\begin{abstract} 
Functional Near-Infrared Spectroscopy (fNIRS) is a non-invasive, real-time method for monitoring brain activity by measuring hemodynamic responses in the cerebral cortex. However, existing systems are expensive, bulky, and limited to clinical or research environments. This paper introduces OpenNIRScap, an open-source, low-cost, and wearable fNIRS system designed to make real-time brain monitoring more accessible in everyday environments. The device features 24 custom-designed sensor boards with dual-wavelength light emitters and photodiode detectors, a central electrical control unit (ECU) with analog multiplexing, and a real-time data processing pipeline. Bench validation and pilot tests on volunteers have confirmed the ability of the system to capture cognitively evoked hemodynamic responses, supporting its potential as an affordable tool for cognitive monitoring and portable neurotechnology applications. The hardware, software, and graphical user interface have all been open-sourced and made publicly available at the following link: \href{https://github.com/tonykim07/fNIRS}{https://github.com/tonykim07/fNIRS}.

\end{abstract}

\section{Introduction}

\begin{figure}[!ht]
\centering
\vspace{-2mm}
\includegraphics[width=1\columnwidth]{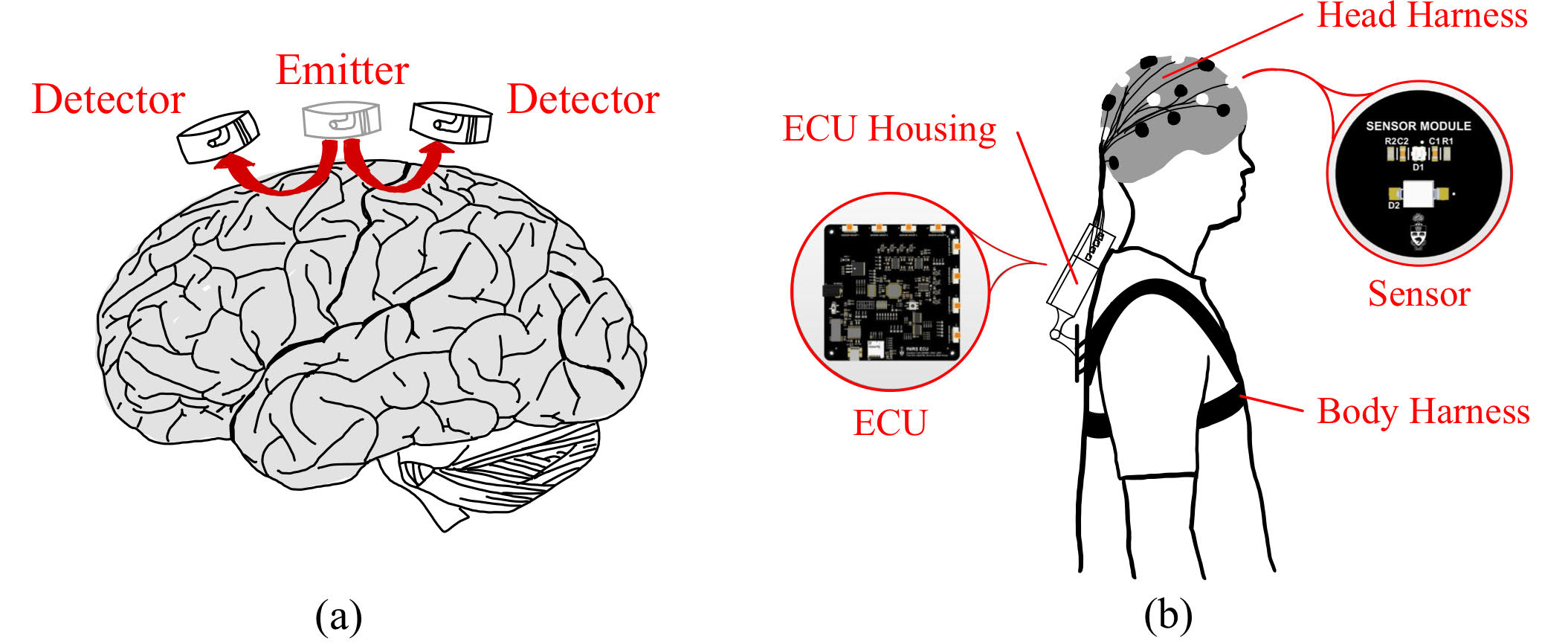}
%add sketch of person (side by side) + new description
\vspace{-5mm}
\caption{(a) Illustration of a typical fNIRS setup with source emitters and detectors placed on the scalp to measure brain activity via light absorption changes. (b) Illustration of the assembled fNIRS-based brain interface system.}
\vspace{-3mm}
\label{fig_intro}
\end{figure}

Monitoring brain activity is critical to diagnosing neurological disorders and enhancing our understanding of human cognitive functions~\cite{zhang2020electronic}. Functional Near-Infrared Spectroscopy (fNIRS) is a non-invasive, real-time neuroimaging technology that measures hemodynamic responses—specifically, changes in oxyhemoglobin (HbO) and deoxyhemoglobin (HbR) concentrations—within the cerebral cortex. As illustrated in Fig.~\ref{fig_intro}(a), fNIRS systems typically consist of light emitters and nearby photodetectors positioned on the scalp to measure changes in light absorption through cortical tissue. Near-infrared light is emitted into the scalp and detected at adjacent points after passing through the brain’s surface layers; changes in absorption are then used to infer underlying neural activity~\cite{intro1}. These measurements are based on the modified Beer-Lambert law, which relates the absorption of near-infrared light to the concentration of the absorbing substance and the distance which light travels through biological media~\cite{Baker:14}.

Beyond simply measuring and processing light data, current fNIRS solutions incorporate hybrid or multimodal brain activity monitoring methods, such as the ability to perform electroencephalogram (EEG) alongside fNIRS~\cite{intro3}. While fNIRS has shown promise in clinical and research settings, many existing devices are expensive, bulky, and designed for specialized use~\cite{intro4}. Although some low-cost alternatives have emerged, they typically provide limited spatial coverage, reducing their potential for comprehensive monitoring~\cite{TSOW2021e00204}.

In this paper, we address these limitations by presenting an open-source, low-cost, and ergonomic fNIRS device OpenNIRScap, as shown in Fig.~\ref{fig_intro}(b), suitable for more inclusive applications in education and personal health monitoring, while maintaining high accuracy and usability. The final system architecture consists of custom-designed sensor modules-24 dedicated photodiode detectors and 8 dual-wavelength source emitters of 660$\,\mathrm{nm}$ and 940$\,\mathrm{nm}$. At the heart of the device is a custom-designed electrical control unit (ECU) built around a low-power microcontroller (MCU) that interfaces with a computer analysis platform. Mechanical supports were also tailored to optimize consistent skin contact and user comfort during operation. This research presents an open-source and low-cost wearable fNIRS device, holding promise for broader applications in cognitive monitoring and everyday neurotechnology.

\begin{figure*}[!ht]
\centering
\includegraphics[width=2\columnwidth]{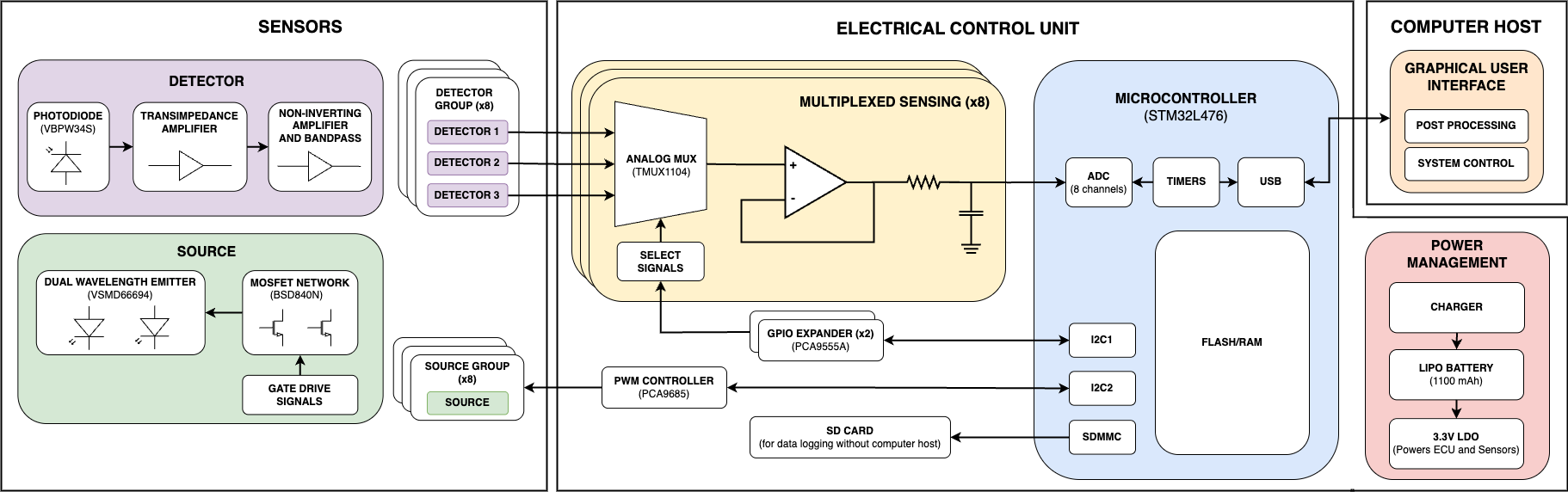}
\caption{A high-level block diagram of the developed system, which consists of light source and detectors sensor modules, an electrical control unit centralized around a microcontroller, and a computer host for data processing and display.}
\vspace{-0mm}
\label{fig_system}
\end{figure*}

\section{System Design}

\subsection{System Architecture} 
A high-level block diagram of the system is shown in Fig.~\ref{fig_system}. The system consists of light source and detector sensor modules using electro-optical components, an fNIRS ECU, a battery to power the electronics, and a computer host with a graphical user interface (GUI) that provides real-time data analysis and device configuration. The key blocks of the fNIRS ECU include (1) analog multiplexed sensing to support a large number of sensor channels, (2) a low-power MCU for emitter control, data acquisition, and device communication between the computer host, and (3) power management including battery charging capabilities. The device supports up to 24 detector channel readings and 8 dual-wavelength light source emitters. Design details are discussed in the following subsections and the implementation is available in the GitHub repository \href{https://github.com/tonykim07/fNIRS}{https://github.com/tonykim07/fNIRS}~\cite{kim2025fnirs}. 

\subsection{Sensor Design} 
fNIRS sensors are required to read changes in HbO and HbR to indicate brain activity. In this work, the sensor design consists of near-infrared light emission and detection circuitry to quantify hemodynamic responses with the modified Beer-Lambert law~\cite{intro1}. 

For emission, the design uses a dual-wavelength LED package (VSMD66694, Vishay Semiconductor Opto Division) of 940$\,\mathrm{nm}$ and 660$\,\mathrm{nm}$ sources. A minimum of two wavelengths is required to differentiate between HbO and HbR in the brain, one above and one below the isosbestic point~\cite{dempsey2015wavelength}. Each wavelength is individually controlled by low-side n-channel MOSFETs (BSD840N, Infineon Technologies) with appropriate in-series resistors to limit the current.

\begin{figure}[!ht]
\centering \vspace{0mm}
\includegraphics[width=\columnwidth]{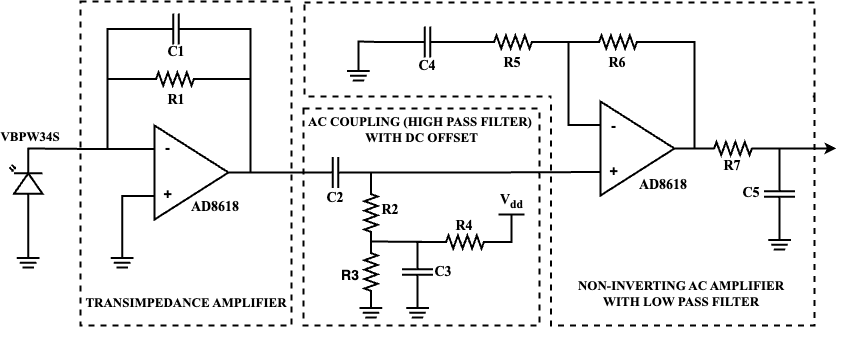}\vspace{0mm}
\caption{Simplified circuit schematic of the detector design.}\vspace{0mm}
\label{fig_detector}
\end{figure}

The simplified detector design is shown in Fig.~\ref{fig_detector}. The detectors use a silicon PIN photodiode (VBPW34S, Vishay Semiconductor Opto Division) with a spectral sensitivity bandwidth of 440$\,\mathrm{nm}$-1100$\,\mathrm{nm}$ and a dark current specified to be 2$\,\mathrm{nA}$ at room temperature. The photodiode is followed by two amplification stages, transimpedance and non-inverting AC gain, implemented on a two circuit operational amplifier (AD8618, Analog Devices Inc.) with a low offset in the range of 23$\,\mu\mathrm{V}$ to 60$\,\mu\mathrm{V}$, a low noise level of 10$\,\mathrm{nV}/\sqrt{\mathrm{Hz}}$, and a low input bias current within 0.2$\,\mathrm{pA}$ to 1$\,\mathrm{pA}$. The bandwidth of the transimpedance amplifier is approximately 13.175$\,\mathrm{kHz}$ set by $R_{1}$ and $C_{1}$, which are feedback resistors and capacitors of values 60.4$\,\mathrm{k \Omega}$ and 200$\,\mathrm{pF}$ respectively. $R_{1}$ is also used to set the gain.

The AC component of the acquired signal contains information about HbO and HbR concentrations~\cite{chenier2007new}, which it is extracted with a high pass filter with a cutoff frequency of 0.0796$\,\mathrm{Hz}$. A DC offset is applied to shift the entire waveform within the available output swing. The gain of the non-inverting amplifier is set by $R_{5}$ and $R_{6}$ where it was empirically chosen to be 101$\,\mathrm{V}/{\mathrm{V}}$ for this design, and an additional low-pass filter is added at the output to increase the stability of the signal reading. The sensors are powered with 3.3$\,\mathrm{V}$ supplied through the fNIRS ECU.

\subsection{Electrical Control Unit Design} 
The fNIRS ECU is the core of the system, providing sensor data acquisition, control, and device interface using a microcontroller (STM32L476, STMicroelectronics) as the main computational unit. Communication to the computer host is achieved with the use of the MCU's USB peripheral with added isolation (ADUM4160, Analog Devices Inc.), and an SD card interface with SDMMC is supported for non-volatile data logging. The fNIRS ECU is powered through a low-dropout regulator downstream of a 1100$\,\mathrm{mAh}$ lithium-ion polymer battery, allowing the device to run for more than 5 hours on a single charge. The fNIRS ECU also incorporates a 2$\,\mathrm{W}$ charger with power available when the computer host is connected.

To account for a large number of sensor channel readings, analog multiplexing is implemented to reduce the number of analog sensing circuits and analog-to-digital converter (ADC) channels on the microcontroller, saving overall cost. A total of 8 precision analog multiplexers (TMUX1104, Texas Instruments) are used, each switching between three sensor channels to a single output. The multiplexers' outputs are followed by voltage follower circuits and low-pass filtering before the MCU's ADC channels. Multiplexer channels are selected with I/O's, where I2C interface I/O expander integrated circuits (PCA9685, NXP USA Inc.) are used, allowing a cheap and small MCU.

Light source emitters are controlled with an I2C (supporting fast mode plus) interface PWM controller (PCA9685, NXP USA Inc.) that drives a total of 16 logic-level PWM channels. The PWMs have a resolution of 12 bits and are configurable to frequencies between 24$\,\mathrm{Hz}$ and 1526$\,\mathrm{Hz}$ and duty cycles and phase shifts between 0\% and 100\%. By default, duty cycles of 100\% are used.

\subsection{Microcontroller System}
The firmware running on the fNIRS ECU's MCU uses hardware timer modules to trigger high-priority data acquisition and light source control while employing low-priority system control in an event loop. The data acquisition pipeline consists of the MCU's ADC sampling and data logging over serial communication, each having its own timer module to control the sampling and logging rate, respectively. The ADC makes use of a direct memory access (DMA) controller to simultaneously capture 8 channel readings with 12-bit resolution. Sampling is also synchronized with analog multiplexing so that memory always has valid data for all sensor readings. Data logging is run in an interrupt service routine, where all sensor data, 24 channel readings, are filtered with a discrete first-order low-pass software filter for tunable noise rejection. The low-pass filter is implemented in the firmware according to the equation:
 \begin{equation}
      y[n] = \alpha \cdot x[n] \cdot (1 - \alpha) * y[n-1]
 \end{equation}
where $y[n]$ is the current output, $x[n]$ is the input from the ADC, $y[n-1]$ is the previous output, and $\alpha$ is the smoothing factor which is:
 \begin{equation}
      \alpha = 1 - e^{-2\pi \frac{f_c}{f_s}}
 \end{equation}
where $f_c$ and $f_s$ are the cutoff and sampling frequencies. The filtered data are populated into a single data buffer and transmitted via USB. By default, the system triggers the ADC start of conversion every 5$\,\mathrm{kHz}$ and data logging is run every 1$\,\mathrm{kHz}$ such that the effective system sampling rate seen by the computer host is 1$\,\mathrm{kHz}$. The analog multiplexers periodically switch between channels every few milliseconds. 

Source emitter control is achieved by writing to the PWM controller registers via I2C to update the duty cycle, phase shift, and frequency of the gate drive signals. The firmware periodically switches between 940$\,\mathrm{nm}$ and 660$\,\mathrm{nm}$ wavelengths, where accurate timing is achieved using a hardware timer module. Emitter control can also be done by the user manually using the GUI, which overrides the default switching pattern, as desired.

\subsection{Computer Host System}
\begin{figure}[!ht]
\centering \vspace{0mm}
\includegraphics[width=\columnwidth]{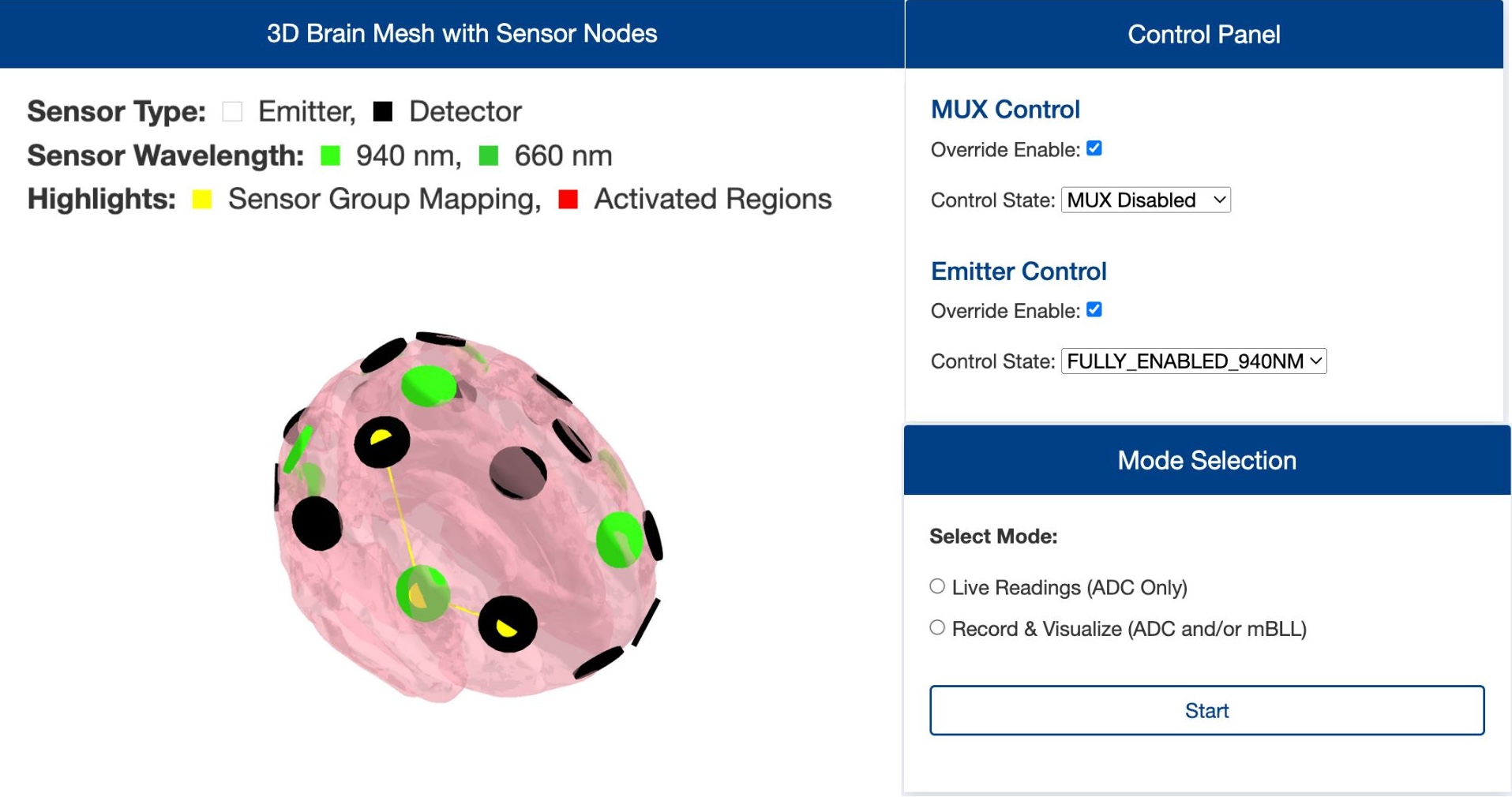}\vspace{0mm}
\caption{Web-based GUI showing sensor mapping, system controls, and mode selection for live or post-processed data visualization.}\vspace{0mm}
\label{GUI}
\end{figure}

To support real-time interaction, the host system includes a web-based GUI powered by a Flask-SocketIO backend for live data streaming. The interface displays raw ADC as well as HbO and HbR concentration changes across channels using a 3D brain mesh and time-series plots, built with Plotly.js, PyQtGraph, and anatomical datasets from BrainNet Viewer~\cite{brainnetviewer}. Users can adjust sensor configurations such as emitter states and multiplexer settings via control panels, with immediate feedback shown in the GUI, as shown in Fig.~\ref{GUI}. All adjustments are transmitted to the hardware using pySerial for synchronized operation.

Underneath the web-based GUI is the software analysis pipeline, which processes buffered data transmitted from the fNIRS ECU by interleaving 660$\,\mathrm{nm}$ and 940$\,\mathrm{nm}$ wavelength samples corresponding to the same sensor sets. This enables the transformation of raw hardware readings into physiologically meaningful metrics.

Owing to the differential absorption properties of HbO and HbR, where the former absorbs more of the 940$\,\mathrm{nm}$ light and transmits more of the 660$\,\mathrm{nm}$ light, and the latter exhibits the inverse behavior, their varying presence modulates the intensity of light received by each detector. Consequently, the raw data buffers compile to a discernible pulsatile waveform over time, whose peak-to-peak intervals correspond directly to instantaneous heart rate.

For hemodynamic analysis, the recorded intensity signals are first converted into changes in optical density. A zero-phase Butterworth band-pass filter (0.01–0.5$\,\mathrm{Hz}$) is applied to retain frequency components associated with cortical hemodynamics while suppressing low-frequency drifts and high-frequency physiological artifacts\cite{10.3389/fnhum.2018.00505}. The filtered signals are subsequently analyzed using the Modified Beer–Lambert Law\cite{Baker:14}, as implemented with the support of the NIRSimple library\cite{10.3389/fnrgo.2023.994969}. Differential pathlength factors are applied according to age and wavelength-specific values proposed by Scholkmann and Wolf\cite{lens.org/064-854-027-896-005}, while molar extinction coefficients are drawn from the reference data provided by Wray et al.\cite{WRAY1988184}. The resulting estimates of HbO and HbR concentration changes are then refined through correlation-based signal improvement\cite{CUI20103039}, which exploits the canonical anti-phase relationship between the two chromophores to attenuate shared noise artifacts.

\subsection{Mechanical Assembly}

\begin{figure}[!ht]
\centering
\includegraphics[width=1\columnwidth]{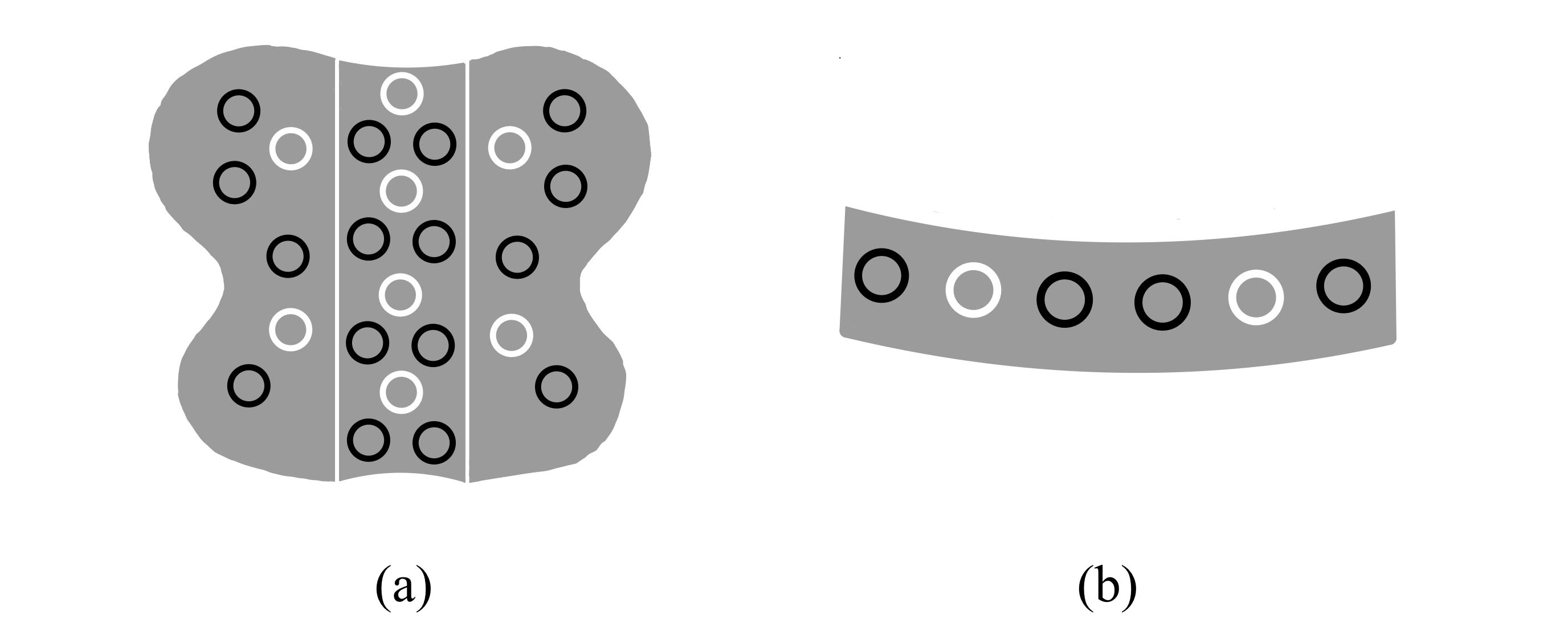}
\vspace{-5mm}
\caption{(a) Full head harness sensor layout. (b) Headband sensor layout.}
\vspace{0mm}
\label{fig_layout}
\end{figure}

Two distinct sensor configurations were implemented to accommodate varying experimental needs: a full-scale head harness layout shown in Fig.~\ref{fig_layout}(a), and a compact headband layout shown in Fig.~\ref{fig_layout}(b). The head harness supports up to 24 sensors strategically distributed across key cortical regions, including the frontal, parietal, occipital, and temporal lobes. Each sensor unit set consists of one module housing a dual-wavelength emitter and a short-channel detector, flanked by two additional modules with long-channel detectors. These triangular sensor sets are arranged with 35$\,\mathrm{mm}$ center-to-center spacing, enabling reliable concurrent short- and long-separation measurements without repositioning\cite{Chance:98}. In contrast, the headband layout is optimized for frontal-lobe monitoring, offering a reduced number of sensor modules in a streamlined arrangement suitable for more lightweight or task-specific applications.

\begin{figure}[!ht]
\centering
\includegraphics[width=0.83\columnwidth]{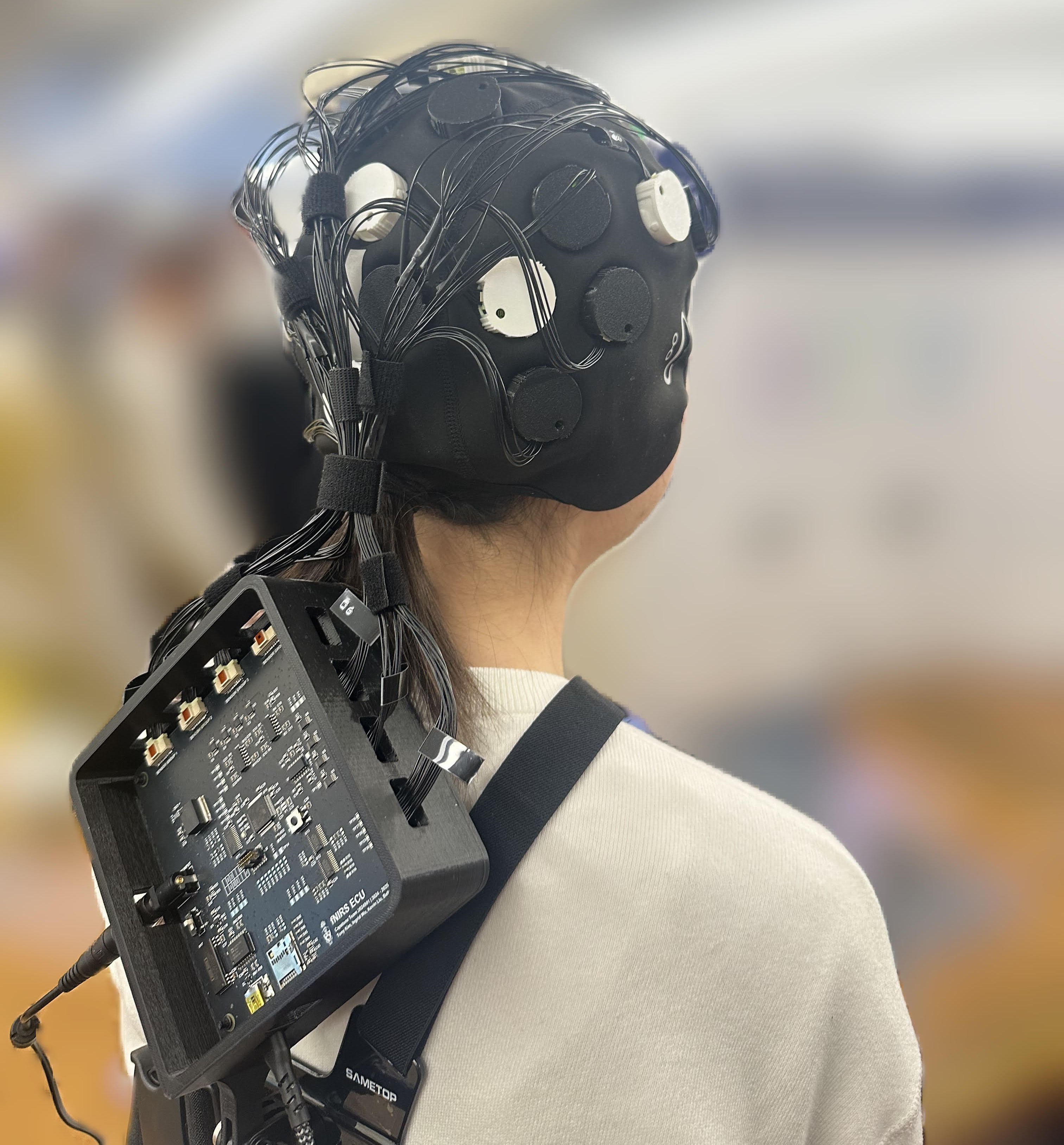}
\vspace{-2mm}
\caption{Fully assembled OpenNIRScap system being worn and in operation.}
\vspace{0mm}
\label{fig_cap_image}
\end{figure}

The remaining mechanical assembly of the test device was designed to ensure stable sensor placement while maintaining participant comfort and mobility. All modules are housed in custom 3D-printed PLA capsules with protective twist-lock caps. These caps apply gentle downward pressure to maintain proper scalp contact, and allow for quick replacement or servicing. Wear tests confirmed that the full device can be donned or removed in less than two minutes and does not significantly restrict head or upper-body movement during seated or ambulatory tasks. An image of the complete system is shown in Fig.~\ref{fig_cap_image}.

\subsection{Cost Breakdown}
The design uses off-the-shelf components assembled onto custom printed circuit boards to achieve a low-cost device. Table \ref{tab_cost_breakdown} outlines the high-level hardware cost breakdown for a fully populated device, rounded to the nearest dollar. A more detailed bill of materials can be found in the design files of the open-source repository.

\begin{table}[htbp]
\caption{Hardware Cost Breakdown}
\begin{center}
\begin{tabular}{|l|c|c|}
\hline
\textbf{Hardware Element} & \textbf{Quantity} & \textbf{Cost (USD)} \\
\hline
Detector Components & 24 & \$175 \\
Emitter Components & 8 & \$12 \\
Sensor PCB Manufacturing & 24 & \$4 \\
fNIRS ECU Components & 1 & \$83 \\
fNIRS ECU PCB Manufacturing & 1 & \$7 \\
Cable Harnesses & 8 & \$73 \\
Rechargeable Battery (1100mAH) & 1 & \$12 \\
PLA Filament & 1 & \$23 \\
Wearable Components & 1 & \$30 \\
\hline
\multicolumn{2}{|r|}{\textbf{Total}} & \textbf{\$419} \\
\hline
\end{tabular}
\label{tab_cost_breakdown}
\end{center}
\end{table}

\section{Experimental Methods and Results}

\subsection{Bench-Level Validation}
\begin{figure}[!ht]
\centering \vspace{0mm}
\includegraphics[width=0.95\columnwidth]{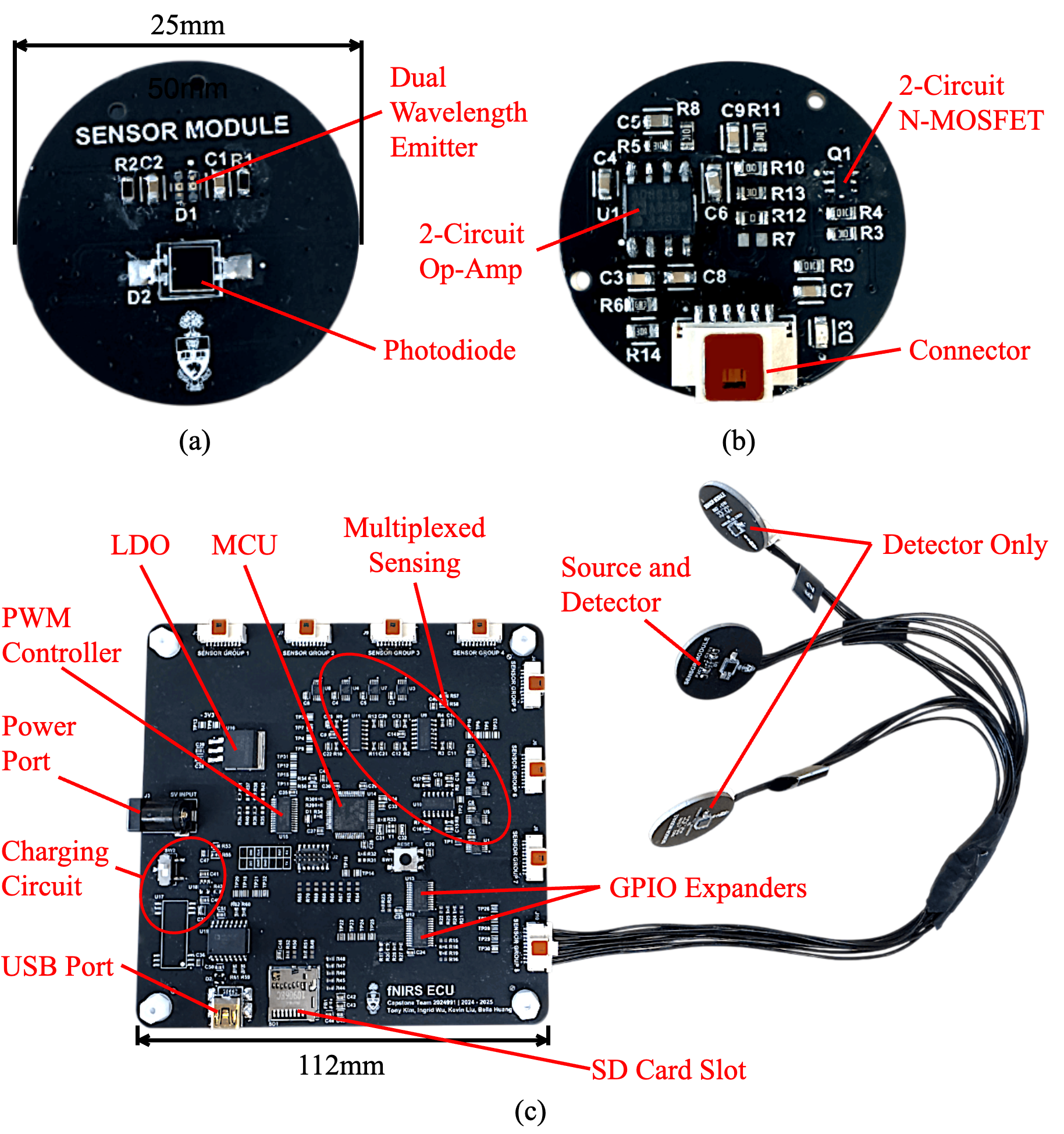}\vspace{0mm}
\caption{(a) Front PCB assembly of the sensor prototype. (b) Back PCB assembly of the sensor prototype. (c) Photo of the assembled fNIRS ECU with one connected sensor group.}
\label{fig_fnirs_assembly}
\end{figure}

The assembled sensors and the fNIRS ECU are shown in Fig.~\ref{fig_fnirs_assembly}. The source and detector circuits are implemented on a singular 25$\,\mathrm{mm}$ diameter circular PCB with the photodiode and dual-wavelength LEDs on the front-facing side. The fNIRS ECU is 112$\,\mathrm{mm}$ by 112$\,\mathrm{mm}$ with 8 connectors for the sensors, each mapping three detectors and one source.

\begin{figure}[!ht]
\centering \vspace{0mm}
\includegraphics[width=1\columnwidth]{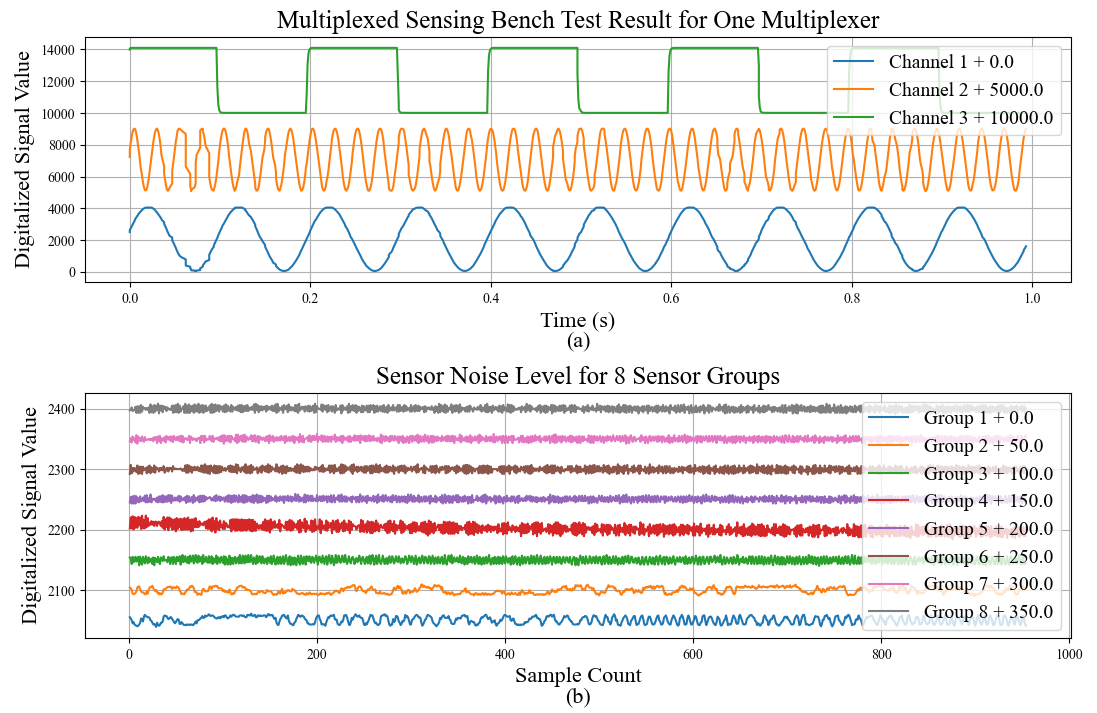}\vspace{-0mm}
\caption{(a) Multiplexed sensing test result using three different waveforms: one 5$\,\mathrm{Hz}$ PWM and two 15$\,\mathrm{Hz}$ and 10$\,\mathrm{Hz}$ sinusoids. (b) Noise level testing collected on each ECU multiplexed group.}
\label{fig_bench_result}
\end{figure}

The device was fully characterized on the bench prior to in vivo experiments. Fig.~\ref{fig_bench_result}(a) shows the verified multiplexed sensing by applying three different waveforms with a functional generator to the input of a given multiplexer on the fNIRS ECU. The acquired readings were calibrated to a precision of $\pm$0.5\% by sweeping the input values across the full range of the MCU's ADC. Fig.~\ref{fig_bench_result}(b) shows noise level tests of the sensors acquired by the fNIRS ECU. The average resulting signal-to-noise ratio (SNR) across all channel readings was 52.302$\,\mathrm{dB}$, with minimum and maximum values being 50.297$\,\mathrm{dB}$ and 53.809$\,\mathrm{dB}$, respectively. 

\begin{figure}[!ht]
\centering \vspace{0mm}
\includegraphics[width=1\columnwidth]{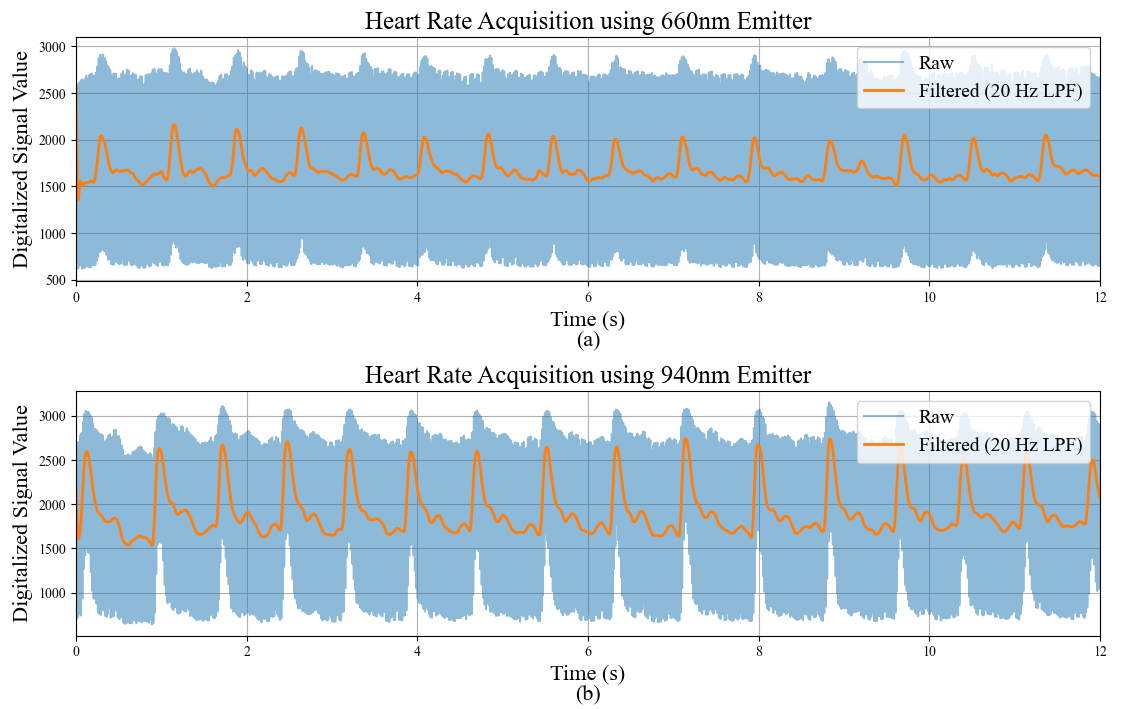}\vspace{-0mm}
\caption{(a) and (b) Heart rate reading results using 660$\,\mathrm{nm}$ and 940$\,\mathrm{nm}$ emitter wavelengths, respectively, demonstrating a 20$\,\mathrm{Hz}$ cutoff low pass software filter.}
\label{fig_heartrate}
\end{figure}

Fig.~\ref{fig_heartrate} (a) and (b) show heart rate signals collected using the sensors placed on the tip of the index finger of a living and healthy participant as preliminary verification of the sensors. All data collected were received by the computer host through the USB interface.

\subsection{Experimental Results}

Experiments were conducted using the headband configuration, with the objective of monitoring hemodynamic responses in the frontal lobe. Three selected healthy participants were seated comfortably in a quiet, controlled environment and instructed to minimize head movement throughout the trial. The protocol consisted of three sequential phases: a 20-second baseline rest period, a 2-minute active task phase involving cognitive and spatial reasoning puzzle-solving, and a final 1-minute post-task rest period. This design was intended to elicit and observe frontal-lobe activation associated with executive function and working memory, followed by recovery dynamics during rest.

\begin{figure}[!ht]
\centering
\includegraphics[width=1\columnwidth]{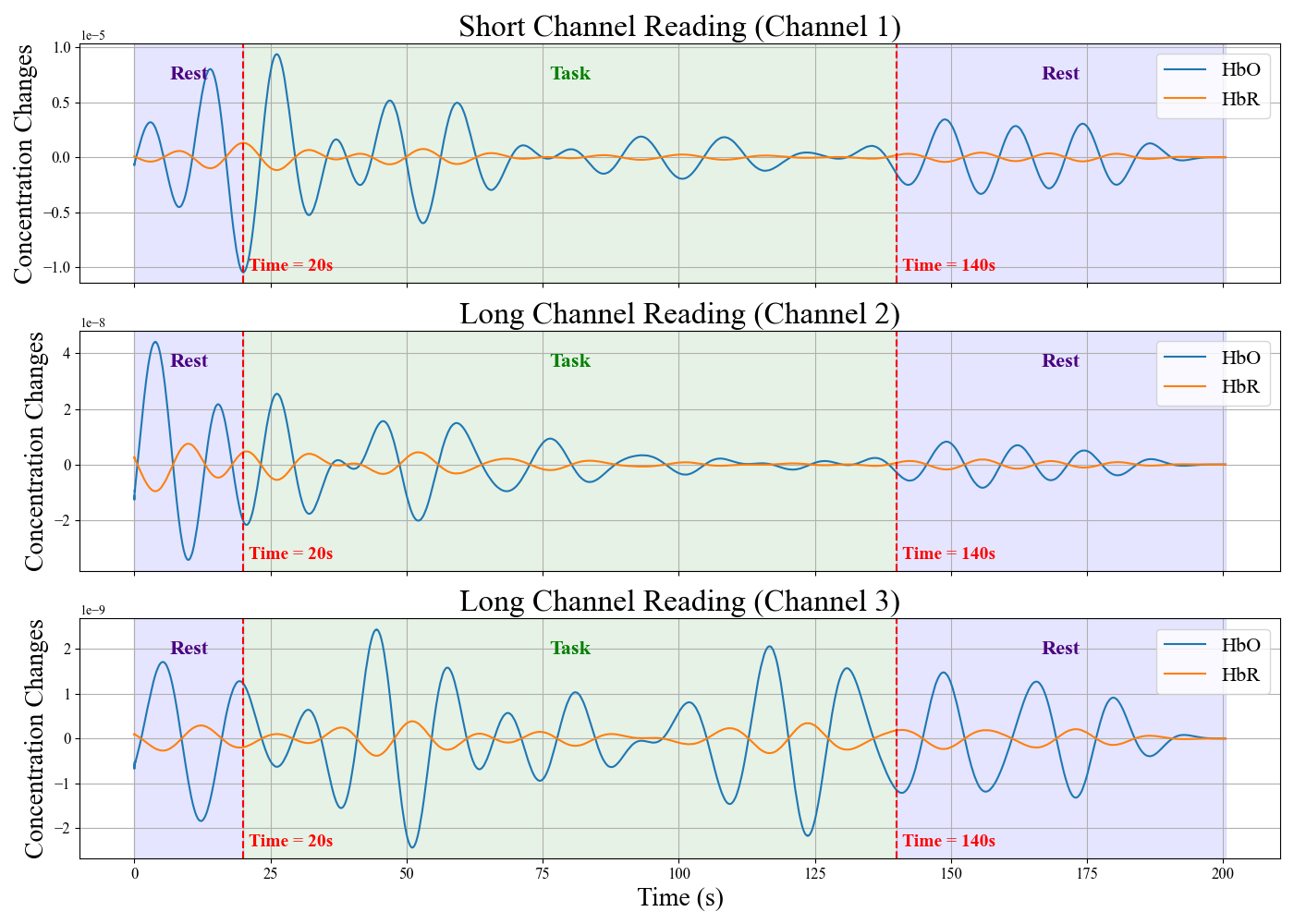}
\vspace{-5mm}
\caption{Plots of hemodynamic readings from a selected sensor set for one of the participants. The sensor set consists of one short channel and two long channel detectors.}
\vspace{0mm}
\label{fig_results}
\end{figure}

Data were collected from all participants. Processed sensor readings were used to extract changes in the molar concentrations of HbO and HbR. Fig.~\ref{fig_results} illustrates a representative result of one of the individuals.

During the active task period, the frontal channels exhibited more frequent and sustained positive $\Delta$HbO accompanied by concomitant negative $\Delta$HbR, signifying increased cerebral oxygen delivery and metabolic demand. These responses appeared to stabilize as the task progressed. Upon task cessation, both signals trended more aggressively during the final rest block, reflecting post-activation recovery of cerebral hemodynamics. Collectively, these findings demonstrate the system’s ability to resolve task-evoked hemodynamic responses and validate the precise temporal correspondence between signal fluctuations and the experimental paradigm.

\section{Conclusion}

This paper presents OpenNIRScap, a compact, low-cost, and fully open-source wearable fNIRS system. The device features 24 custom-designed sensor modules with dual-wavelength light sources and photodiode detectors, coordinated by an analog-multiplexed sensing fNIRS ECU. Despite its affordability—built entirely from off-the-shelf components at a total cost of \$419 USD—the device supports high spatial coverage with significantly more channels than typical low-cost systems. A comparison with existing and comparable fNIRS systems can be found in Table \ref{tab_comparision}.

\begin{table}[htbp]
\caption{FNIRS Device Comparison}
\begin{center}
\begin{tabular}{|l|c|c|c|c|}
\hline
\shortstack{\textbf{Specs}\\\textbf{}} & \shortstack{\textbf{OpenNIRScap}\\\textbf{}} & \shortstack{\\\textbf{DIY-fNIRS}\\\textbf{headband}~\cite{TSOW2021e00204}} & \shortstack{\textbf{biosignalsplux}\\\textbf{fNIRS Sensor}~\cite{pluxbiosignals}}\\
\hline
Coverage & 24 channels & 4 channels & 1 channel\\
Sampling Rate & 1$\,\mathrm{kHz}$ & 10$\,\mathrm{Hz}$ & 500$\,\mathrm{Hz}$\\
Resolution & 12-bit & 12-bit & 16-bit\\
Wearable & Yes & Yes & No\\
Cost (USD) & \$419 & \$215 & \$789\\
Open-Source & Yes & Yes & No\\
\hline
\end{tabular}
\end{center}
\label{tab_comparision}
\end{table}

Featuring a high SNR of 52.302$\,\mathrm{dB}$, a default 1$\,\mathrm{kHz}$ sampling rate, and a feature-rich software platform for real-time data analysis, the device successfully demonstrates its efficacy in resolving task-evoked hemodynamic changes and establishes itself as a reproducible and versatile tool for neurotechnology research, education, and accessible personal use. The developed open-source platform enables seamless integration of additional sensors and software-based signal processing, including on-device machine learning~\cite{sun2023design}, allowing for multimodal brain interfacing~\cite{yang2024multi} across a wide range of applications.

\bibliographystyle{IEEEtran}
\bibliography{ref}

\end{document}